\documentclass[aps,prd,twocolumn,preprintnumbers,floatfix,nofootinbib,notitlepage,showkeys,showpacs]{revtex4-1}

\usepackage[utf8x]{inputenc}

\usepackage{graphicx}
\usepackage{hyperref}
\usepackage{latexsym}
\usepackage{amsmath}
\usepackage{amssymb}
\usepackage{bbm}
\usepackage{ulem}
\usepackage{pdfsync}
\usepackage{epsfig}
\usepackage{epstopdf}
\usepackage{color}
\usepackage{comment}
\usepackage{placeins}
\usepackage{slashed}

\newcommand{\nn}{\nonumber}
\newcommand{\ef}{{\rm eff.}}
\def\lsim{\mathrel{\raise.3ex\hbox{$<$\kern-.75em\lower1ex\hbox{$\sim$}}}}
\def\gsim{\mathrel{\raise.3ex\hbox{$>$\kern-.75em\lower1ex\hbox{$\sim$}}}}

\begin{document}

\title{Phase transition and gravitational wave phenomenology of scalar conformal extensions of the Standard Model}

\author{Luca Marzola}
\email{luca.marzola@cern.ch}
\affiliation{National Institute of Chemical Physics and Biophysics, \\ R\"avala pst.~10, 10143 Tallinn, Estonia}
\author{Antonio Racioppi}
\email{antonio.racioppi@kbfi.ee}
\affiliation{National Institute of Chemical Physics and Biophysics, \\ R\"avala pst.~10, 10143 Tallinn, Estonia}
\author{Ville Vaskonen}
\email{ville.vaskonen@kbfi.ee}
\affiliation{National Institute of Chemical Physics and Biophysics, \\ R\"avala pst.~10, 10143 Tallinn, Estonia}

\begin{abstract}
Thermal corrections in classically conformal models typically induce a strong first-order electroweak phase transition, thereby resulting in a stochastic gravitational wave background that could be detectable at gravitational wave observatories. After reviewing the basics of classically conformal scenarios, in this paper we investigate the phase transition dynamics in a thermal environment and the related gravitational wave phenomenology within the framework of scalar conformal extensions of the Standard Model. We find that minimal extensions involving only one additional scalar field struggle to reproduce the correct phase transition dynamics once thermal corrections are accounted for. Next-to-minimal models, instead, yield the desired electroweak symmetry breaking and typically result in a very strong gravitational wave signal.
\end{abstract}

\maketitle

\section{Introduction} 
\label{sec:Introduction}

The LIGO collaboration has recently observed for the first time the direct effects of gravitational waves on matter~\cite{Abbott:2016blz}, marking the beginning of gravitational wave astronomy. This new experimental field pursues the fascinating possibility that important information about the evolution of the Universe could be encoded in a gravitational wave background. In light of this, as the boundaries of the observed gravitational wave spectrum are to be considerably extended by forthcoming space-based interferometers such as LISA~\cite{Baker:2007}, we can expect that gravitational wave astronomy will soon produce new important observables and benchmarks for models of particle physics and gravitation. In this regard, the possible occurrence of phase transitions in the evolution of our Universe is one of the matters within the reach of current and next-generation gravitational waves experiments. 

The dynamics of phase transitions have been investigated in connection to a number of different topics, ranging from the problem of the baryon asymmetry of the Universe~\cite{No:2011fi,Huang:2016odd,Chala:2016ykx,Katz:2016adq,Artymowski:2016tme,Vaskonen:2016yiu,Dorsch:2016nrg,Beniwal:2017eik,Kobakhidze:2017mru,Saeedhosini:2017dsh} to the natural appearance of such phenomena in several high energy completions of the Standard Model (SM)~\cite{Schwaller:2015tja,Kakizaki:2015wua,Jinno:2015doa,Huber:2015znp,Leitao:2015fmj,Jaeckel:2016jlh,Dev:2016feu,Jinno:2016knw,Huang:2016cjm,Hashino:2016xoj,Kubo:2016kpb,Balazs:2016tbi,Huang:2017laj,Baldes:2017rcu,Chao:2017vrq,Huang:2017laj,Lewis:2017dme,Tsumura:2017knk}. Phase transitions which directly or indirectly result in the generation of the electroweak scale have also been analysed within the framework of classically conformal (or scale invariant) models~\cite{Espinosa:2007qk,Espinosa:2008kw,Konstandin:2011dr,Konstandin:2011ds,Servant:2014bla,Fuyuto:2015vna,Sannino:2015wka}, often in connection to other open problems of contemporary physics such as the origins of Dark Matter (DM) and the Inflationary dynamics~\cite{Cooper:1982du,Espinosa:2008kw,Foot:2010av,Ishiwata:2011aa,Okada:2012sg,Heikinheimo:2013fta,Hambye:2013sna,Farzinnia:2013pga,Khoze:2013uia,Gabrielli:2013hma,Allison:2014zya,Allison:2014hna,Kannike:2014mia,Kannike:2015apa,Karam:2015jta,Kannike:2015kda,Wang:2015cda,Marzola:2015xbh,Kannike:2016jfs,Karam:2016rsz,Kannike:2016wuy,Marzola:2016xgb}. An important feature of classically conformal models is that the phase transition associated to the insurgence of the electroweak scale is generally of the first order and very strong. It is then plausible that the related dynamics result in sizeable gravitational wave signals, which would be nowadays encoded in a stochastic background.

Attracted by this possibility, in this paper we focus on the classically conformal extensions of the SM expanding on previous analysis~\cite{Espinosa:2007qk,Espinosa:2008kw,Sannino:2015wka} of the topic, accounting for the impact of thermal corrections on the dynamics of phase transition and extending the phenomenology of these models to cover their possible gravitational wave signal. In more detail, we demonstrate in a general way that scalar conformal models have indeed the capability to give rise to gravitational wave signals that current and future dedicated observations could detect. By applying the developed formalism we then find that the simplest scalar conformal extension of the SM is, at best, strongly constrained by the phenomenology associated to the phase transition once thermal corrections and perturbativity arguments are taken into account. The next-to-minimal models that rely on two new scalar fields, instead, bypass the shortcomings affecting the minimal scenario once the phase transition dynamics is relegated to early epochs in the evolution of the Universe. In this case, we find that the scenarios generally give rise to very strong gravitational wave signals detectable at the current and next-generation gravitational wave interferometers.

The paper is organised as follows: in section~\ref{sec:Gravitational signatures of conformal models} we review the computation of the phase transition temperature and of the emitted gravitational wave spectrum for a classically conformal theory with two scalar fields. The results of this simple model are discussed in relation to the expected sensitivities of aLIGO and LISA observatories. In section~\ref{sec:Single scalar extension of the Standard Model} we apply our formalism to a conformal scalar singlet extensions of the SM, extending the work in~\cite{Espinosa:2008kw} to scenarios where the Higgs vacuum expectation value (VEV) is induced by the new scalar via a portal coupling. The next-to-minimal models are addressed in section~\ref{sec:Next-to-minimal models}, whereas our conclusions are presented in section~\ref{sec:conclusions}.

\section{Gravitational wave signatures of conformal models} 
\label{sec:Gravitational signatures of conformal models}

We briefly review here the key steps in the computation of the potential at finite temperature for a conformal scalar field model. At zero temperature, the one-loop scalar potential of $n$ scalar fields $\phi_j$, $j\in\{1,\dots, n\}$, is given by the Colemann-Weinberg result~\cite{Coleman:1973jx} 
\begin{equation}
	\label{eq:CW}
	V =\hspace{-.3cm} \sum_{i,j,k,l = 1}^n \hspace{-.3cm} \lambda_{ijkl} \phi_i \phi_j \phi_k \phi_l + 
	\sum_{k=1}^n \frac{g_k M_k^4}{64\pi^2} \log\left(\frac{M_k^2}{\mu^2}\right) + \delta V  \,,	
\end{equation}
where $\delta V$ contains the counterterms, $\mu$ is the renormalisation scale and $M_k$ and $g_k$ are, respectively, the field dependent tree level mass and the number or intrinsic degrees or freedom of the particle $k$. Notice that in our convention $g_k$ assumes positive values for bosons and negative ones for fermions. 

Consider now a direction in the scalar field space defined by $\phi = \sum_{j=1}^n a_j \phi_j$, where $\sum_{j=1}^n a_j^2 = 1$. Along this direction the tree level mass of the scalar field $\phi$ can be written as $M_k = W_k \phi$, where $W_k$ depends only on adimensional couplings. The potential in eq.~\eqref{eq:CW} along the direction $\phi$ is then written as~\cite{Sannino:2015wka}
\begin{align}
	V &= 
 \frac 14(\lambda_\phi+\delta\lambda_\phi)\phi^4+ A \phi^4 + B \phi^4 \log\left(\frac{\phi}{\mu}\right) \,,
\end{align}
where
\begin{equation}
	A = \sum_{k=1}^n \frac{g_k W_k^2}{64\pi^2} \log{W_k^2} \,, \quad B = \sum_{k=1}^n \frac{g_k W_k^4}{32\pi^2} \,.
\end{equation}
We require that the tree level potential is flat along the direction $\phi$, $\lambda_\phi=0$, and set the counterterm $\delta\lambda$ via the renormalisation condition ${\rm d}^4V/{\rm d}\phi^4|_{\phi=e^{-11/6} v_\phi}=0$. Here $v_\phi$ is the VEV of $\phi$ induced in the spontaneous breaking of the symmetry via the Coleman-Weinberg mechanism. In this way the scalar potential along $\phi$ finally reads
\begin{equation}
	\label{eq:CWpot}
	V = B \phi^4 \left( \log\left(\frac{\phi}{v_\phi}\right) - \frac{1}{4} \right) \,,
\end{equation}
and for the mass of $\phi$ we have $M_\phi^2 = 4Bv_\phi^2$.

The one-loop finite temperature corrections to the above scalar potential are given by~\cite{Laine:2016hma}
\begin{equation}
	V_T = \sum_{k=1}^n J_T(M_k) + \sum_{k\in{\rm bosons}} D_T(M_k,\Pi_k) \,,
\end{equation}
where the thermal integral $J_T$ is specified by
\begin{equation}
	J_T(M_k) =  g_k T \int\frac{{\rm d}^3p}{(2\pi)^3} \log(1\mp e^{-E/T}) \, \label{eq:JT},
\end{equation}
and the upper (lower) sign is for bosons (fermions). The contribution from the re-summed daisy diagrams instead amounts to
\begin{equation}
	D_T(M_k,\Pi_k) = \frac{g_k T}{12\pi} \left(M_k^3 - (M_k^2+\Pi_k(T))^{\frac{3}{2}}\right) \,,
\end{equation}
and depends on the Debye mass $\Pi_k$ of the boson $k$. Notice that for $T\gg M_k$ the thermal integral can be approximated as 
\begin{align}
	J_T(M_k\ll T) &= c_k g_k M_k^2 T^2/12+\text{const.}\,,
\end{align}
where
\begin{equation}
	c_k = 
	\begin{cases}
		1 &\text{ (bosons)}\\
		-1/2 &\text{ (fermions)}\\
	\end{cases}\,.
\end{equation}
In this regime, the one-loop finite temperature effective potential along the $\phi$ direction is then given by
\begin{equation}
	 \label{finiteTpotential}
	V_{\ef} = V + V_T = B \phi^4 \left( \log\left(\frac{\phi}{v_\phi}\right) - \frac{1}{4} \right) + C T^2 \phi^2  \,,
\end{equation}
where we defined $C = \sum_k c_k g_k W_k^2/12\geq0$. Notice that the second derivative of the scalar potential at $\phi=0$ matches $C T^2$, so the thermal potential has necessarily a local minimum at $\phi=0$ independently of the specifics of the underlying conformal model. 

\begin{figure}
	\centering
\includegraphics[width=.42\textwidth]{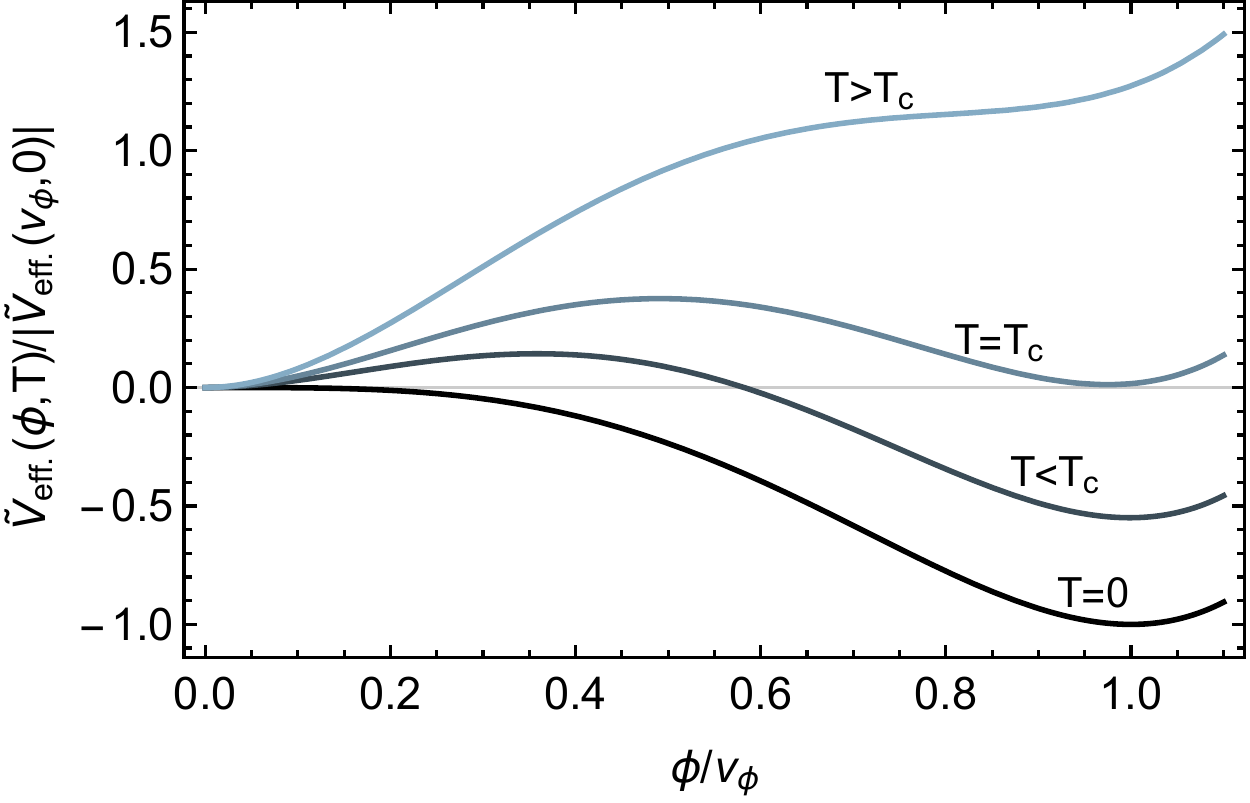}
\caption{The scalar field potential $\tilde{V}_\ef(\phi,T) = V_\ef(\phi,T)-V_\ef(0,T)$ which accounts for the full one-loop thermal integral and the re-summed daisy diagrams in the proposed two-scalars toy model plotted for different temperatures.}
\label{fig:potential}
\end{figure}

For sake of definiteness, consider now a minimal scenario comprising $n=2$ scalar fields, $\phi$ and $\sigma$, with a $\mathbb{Z}_2$ symmetry that bars linear terms in the potential. Suppose also that the scalar field $\phi$, which lies along the flat direction of the tree-level potential, give a mass $M_\sigma^2 = \lambda_p \phi^2/2$ to the field $\sigma$ via a (positive) portal coupling $\lambda_p$, so that $B=\lambda_p^2/128\pi^2$. The one-loop thermal part of the potential is given by 
\begin{equation}
	V_T = J_T(M_\sigma)+D_T(M_\sigma,\Pi_\sigma)\,,
\end{equation}
where $\Pi_\sigma=\lambda_p T^2/3$ and $g_\sigma=1$. We find that using the approximation $V_T = M_\sigma^2 T^2/12$ leads to underestimating the reference phase transition temperatures ($T_c$ and $T_n$, as explained later) by a factor of less than $10$ but does not alter the qualitative discussion of the example at hand. The evolution of the total potential $V_T$ along the $\phi$ direction as a function of temperature is shown in fig.~\ref{fig:potential}. The temperature at which the minima at $\phi=0$ and $\phi\neq 0$ are degenerate is the critical temperature $T_c$.

\subsection*{The phase transition} 
\label{sub:The Phase transition}

As made clear from fig.~\ref{fig:potential}, for temperatures of the thermal bath large enough to lift the minimum of the potential, the field\footnote{Or, rather, its expectation value which obeys the classical equations of motions.} is drawn towards the origin, in a way that $v_\phi=0$ and possible symmetries are restored. The field is stuck at this point until $T<T_c$, when the potential develops again a new global minimum characterised by a non-zero VEV. As mentioned before, however, thermal corrections in conformal models necessarily result in a potential barrier which separates the origin, a local minimum of the potential, from the minimum corresponding to the true vacuum of the theory. Because such potential barrier disappears only for $T=0$, the phase transition from $v_\phi=0$ to $v_\phi\neq0$ in conformal models is always of first order for any finite temperature. We remark that this is a model independent result which applies to all classically conformal scenarios, including the conformal extensions of the Standard Model. Consequently to the abrupt transition, the phase transition proceeds via nucleation and consequent expansion of bubbles inside of which the field is in the broken phase of the theory. The bubble nucleation rate per unit of time and volume is given by~\cite{Linde:1981zj}
\begin{equation}
\Gamma(T) \simeq T^4\left(\frac{S_3}{2\pi T}\right)^{\frac{3}{2}}\exp\left( -\frac{S_3}{T} \right) \,,	
\end{equation}
where
\begin{equation}
	 \label{fullS3}
	S_3 = 4\pi \int r^2 {\rm d}r \left( \frac{1}{2}\left(\frac{{\rm d}\phi}{{\rm d}r}\right)^2 + V_\ef(\phi,T) \right)
\end{equation}
is the three-dimensional Euclidean action for an O(3)-symmetric bubble\footnote{We remark that the action is  minimised by an O(3)-symmetric solution rather than an O(4) one. This is expected whenever the potential barrier arises purely from thermal effects.}. The largest contribution to the bubble nucleation rate arises from the path (in the field space) which minimizes $S_3$, obtained by solving the equation
\begin{equation}
	\frac{{\rm d}^2 \phi}{{\rm d} r^2} + \frac{2}{r} \frac{{\rm d} \phi}{{\rm d} r} = \frac{{\rm d} V_\ef}{{\rm d} \phi}
\end{equation}
with boundary conditions ${\rm d}\phi/{\rm d}r=0$ at $r=0$, and $\phi\to0$ at $r\to\infty$. The bubble nucleation temperature, $T_n$, is then defined as the temperature at which the probability of producing at least one bubble per horizon volume in Hubble time approaches the unity:
\begin{equation}
	\frac{4\pi}{3} \frac{\Gamma(T_n)}{H(T_n)^4} \simeq 1 \,.
\end{equation}

\begin{figure}
\centering
\includegraphics[width=.45\textwidth]{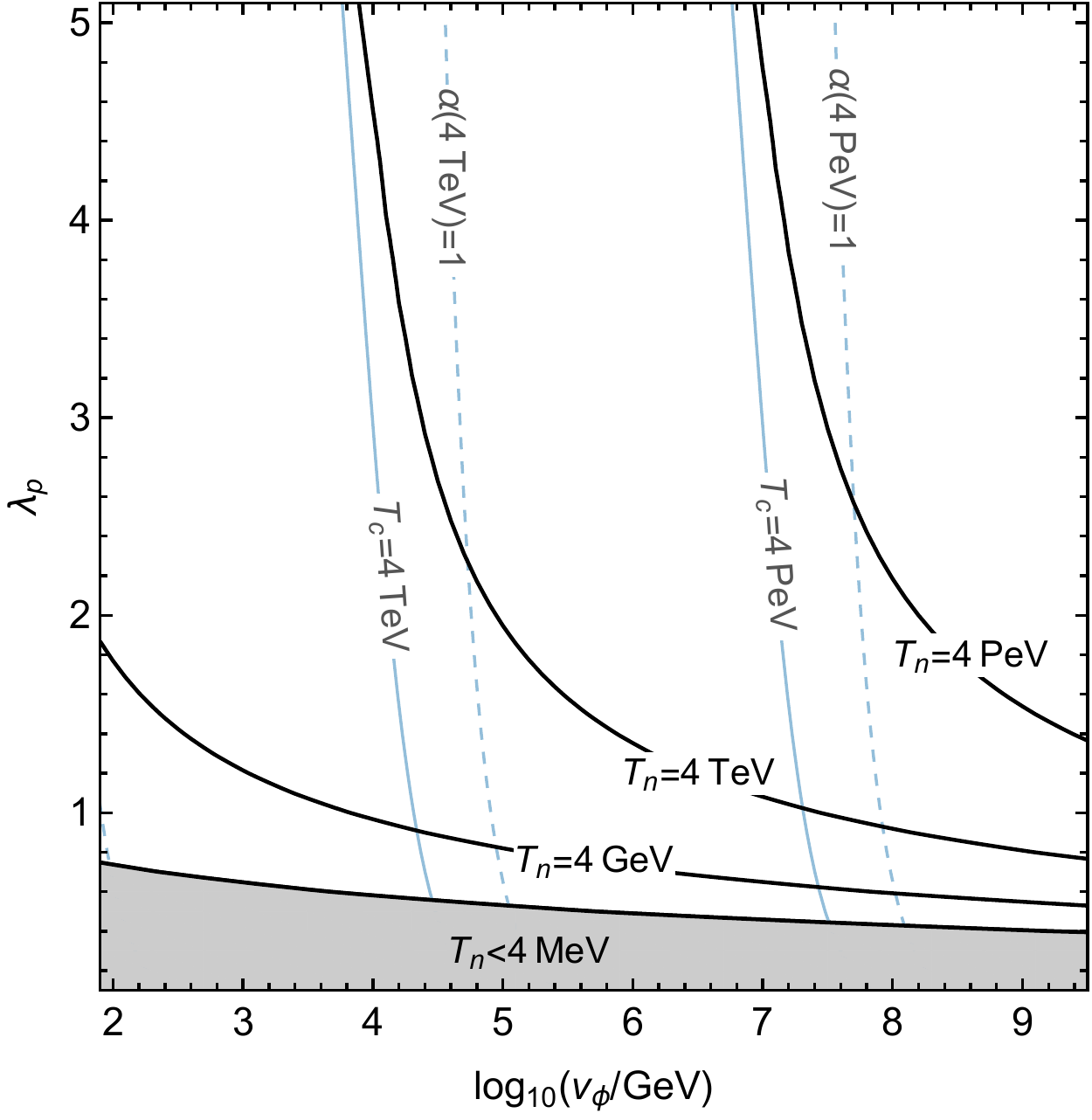}
\caption{The phase transition in classically conformal models. The thick solid lines are the isocontours of the phase transition temperature $T_n$, the thin solid lines are instead those of the critical temperature $T_c$. The dashed lines show the temperature corresponding to $\alpha(T)=1$. The grey region is excluded by the requirement that the phase transition occur at temperatures above the BBN one.}
\label{fig:transition}
\end{figure}

We plot in fig.~\ref{fig:transition} the bubble nucleation temperature as a function of the portal coupling $\lambda_p$ and the VEV $v_\phi$. Requiring that the phase transition occurs above the big bang nucleosynthesis (BBN) temperature, $T_{\rm BBN} \simeq 4$\,MeV, excludes values of the portal coupling $\lambda_p\lsim0.5$. Notice also that for $\lambda_p\lsim 2$ the bubble nucleation temperature is much lower than the critical temperature $T_c$, implying a large amount of supercooling. The transition is then very strong and we can consequently expect a sizeable gravitational wave signal.

\subsection*{Gravitational wave production} 
\label{ssub:Gravitational wave production}

\begin{figure}
	\centering
\includegraphics[width=.42\textwidth]{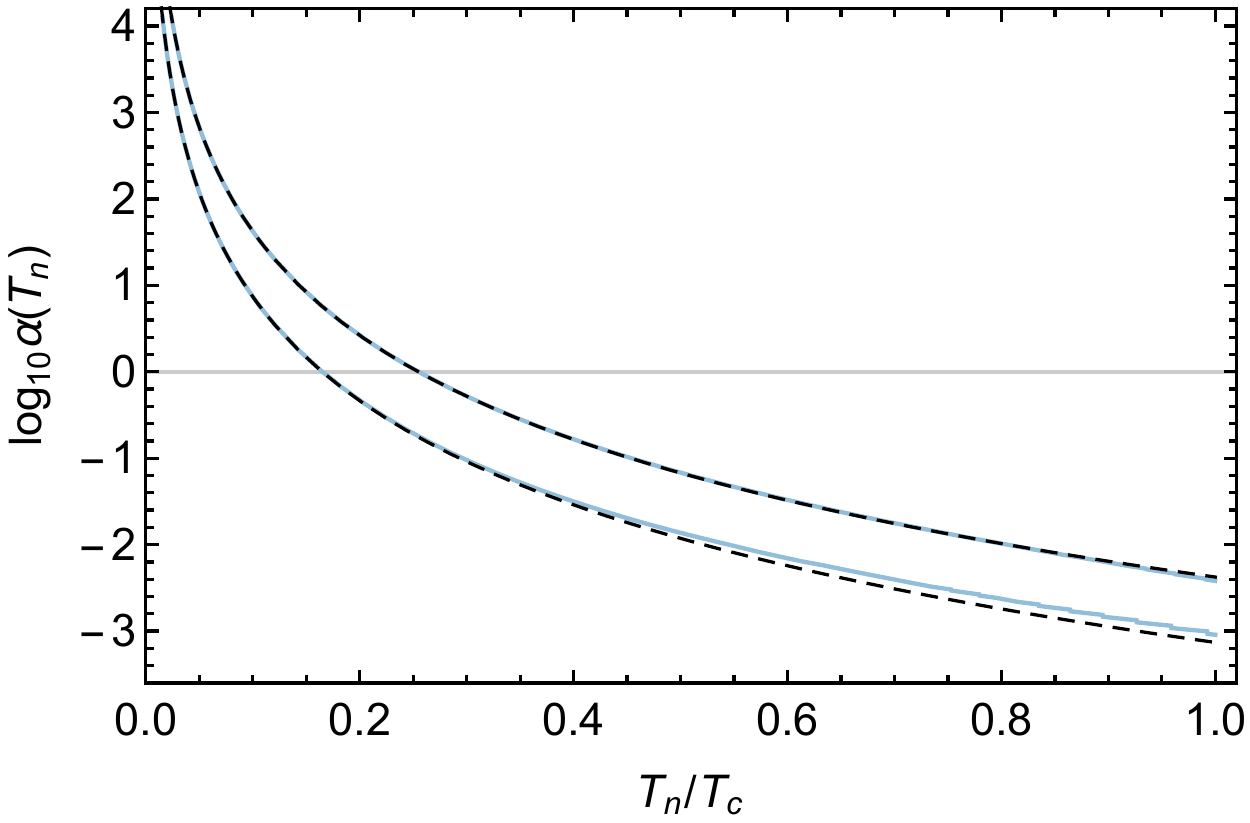}
\caption{Plot of $\alpha$ as a function of $T_n/T_c$ for $\lambda_p=1$ (top lines) and $\lambda_p=5$ (bottom lines). The solid lines correspond to the result obtained with eq.~\eqref{eq:alphaexact} whereas the dashed lines are for the approximation in eq.~\eqref{eq:alphaapp}.}
\label{alpha}
\end{figure}

The ratio of the vacuum energy released during the phase transition to the energy density of the radiation bath at a temperature $T$ is given by~\cite{Espinosa:2010hh}
\begin{equation}
	\label{eq:alphaexact}
	\alpha(T) = \frac{1}{\rho_\gamma}\left(\Delta V_\ef - \frac{T}{4} \Delta \left(\frac{{\rm d}V_\ef}{{\rm d}T}\right) \right) \,,
\end{equation}
where we indicated with $\Delta X$ the difference $X(0)-X(v_\phi)$ for a quantity $X$. As shown in fig.~\ref{alpha}, the value of $\alpha$ for $T\ll T_c$ can be very well approximated by 
\begin{equation}
	\label{eq:alphaapp}
	\alpha(T) \simeq \frac{ V(0)-V(v_\phi)} {\rho_\gamma(T)} \simeq 6\times 10^{-4}\frac{\lambda_p^2 v_\phi^4}{g_* T^4}\,,
\end{equation}
where $g_*$ is the effective number of relativistic degrees of freedom in the thermal plasma. Given that $T_c\propto v_\phi$, the curves shown in fig.~\ref{alpha} do not depend on the value of $v_\phi$. 

The thin dashed lines in fig.~\ref{fig:transition} denote instead the configurations of the model for which $\alpha(T)\simeq1$, with increasingly larger values falling on the right hand side of the lines. We see that for small $\lambda_p\lsim 2$ the vacuum energy released in the transition is much larger than the energy density in the radiation bath, $\alpha\gg1$, as suggested by the large hierarchy between critical and nucleation temperature. In this regime we then expect that in classically conformal models the dynamics of phase transition result in a substantial reheating of the system, consequent to the scalar field tunnelling through the thermal potential barrier~\cite{Megevand:2016lpr}. 

\begin{figure}[h]
	\centering
\includegraphics[width=.45\textwidth]{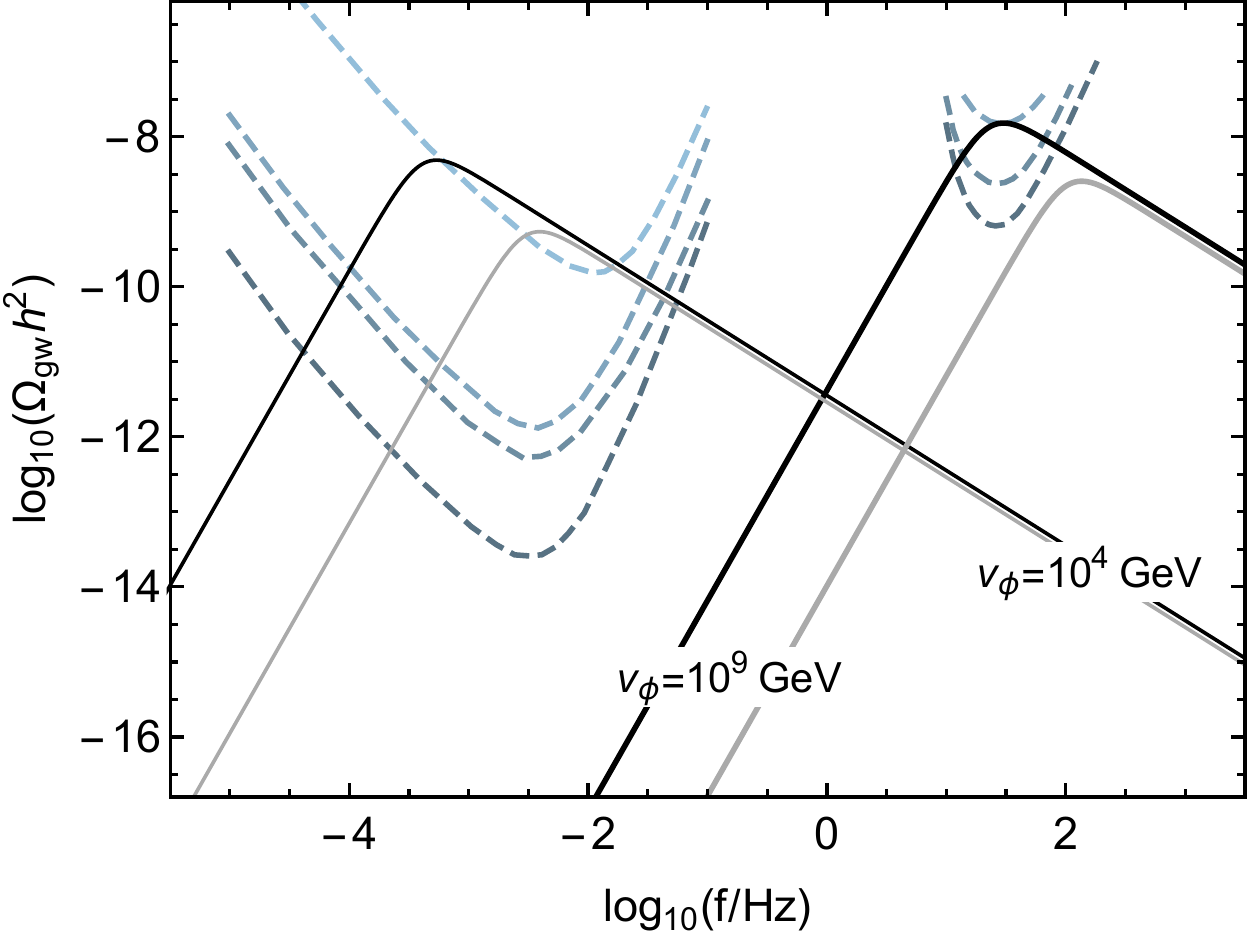}
\caption{The gravitational wave spectrum obtained for the considered model assuming the values of the relative parameters reported in table~\ref{bench}. The black and grey solid lines show the gravitational wave spectra resulting from the phase transition dynamics for $\lambda_p=1$ and $\lambda_p=2$, respectively. The dashed lines correspond instead to the expected sensitivities for different configurations of the LISA detector (low frequency region)~\cite{Caprini:2015zlo} and the reach of the LIGO experiment after several phases of running (high frequency region)~\cite{TheLIGOScientific:2016wyq}.}
\label{fig:gwspectrum}
\end{figure}

For these values of the parameters, it can be shown that plasma effects do not play an important role in the bubble expansion~\cite{Caprini:2015zlo}. Then, as the bubble wall velocity approaches the speed of light, the gravitational wave signal arises purely from the scalar field contribution~\cite{Kamionkowski:1993fg,Caprini:2015zlo}, yielding:
\begin{equation}
	\Omega_{\rm gw}h^2 = \frac{4.9\times 10^{-6}\left(\frac{f}{f_{\rm env}}\right)^{2.8}}{1+2.8\left(\frac{f}{f_{\rm env}}\right)^{3.8}} \left( \frac{H_*}{\beta} \right)^2 \left( \frac{100}{g_*} \right)^{\frac{1}{3}} \,.
\end{equation}
Here $H_*$ is the value of the Hubble parameter at $T=T_*$, corresponding to the temperature of the radiation bath after the phase transition, whereas $f_{\rm env}$ is the redshifted peak frequency of the spectrum as measurable today,
\begin{equation}
	\frac{f_{\rm env}}{\rm Hz} = 3.5\times 10^{-6} \left( \frac{\beta}{H_*} \right) \left( \frac{T_*}{100\,{\rm GeV}} \right) \left( \frac{g_*}{100} \right)^{\frac{1}{6}} \,.
\end{equation}
The parameter $\beta$ describes instead the duration of the transition,
\begin{equation}
	\frac{\beta}{H_*} = T_n \left.\frac{{\rm d}}{{\rm d} T} \frac{S_3}{T}\right|_{T=T_n} \,.
\end{equation}
In fig.~\ref{fig:gwspectrum} we compare the sensitivity of current and future generation of gravitational waves observatories with the gravitational wave spectrum obtained in the considered general model for the benchmark points given in table~\ref{bench}. As we can see, these experiments have the capability to detect the gravitational echoes of the phase transition in classically conformal models and the results we obtain demonstrate the potential impact of gravitational wave phenomenology on these scenarios.

\begin{table}[h]
\begin{tabular}{c c c c c c}
	$\lambda_p$ & $v_\phi/$GeV & $T_n/$GeV & $T_*/$GeV & $T_c/$GeV & $\beta/H_*$\tabularnewline 
	\hline
$1$ & $10^4$ & $5.68$ & $493$ & $1940$ & $23.5$ \\
$2$ & $10^4$ & $512$ & $1200$ & $2990$ & $70.1$ \\
$1$ & $10^9$ & $6.76\times10^4$ & $4.88\times10^7$ & $1.94\times10^8$ & $13.6$ \\
$2$ & $10^9$ & $2.11\times10^7$ & $9.00\times10^7$ & $2.99\times10^8$ & $32.5$ \tabularnewline
\hline
\end{tabular}
\caption{The values of the parameters adopted in the plot of fig.~\ref{fig:gwspectrum}.}
\label{bench}
\end{table}

\section{Minimal conformal extension of the Standard Model } 
\label{sec:Single scalar extension of the Standard Model}


We apply now the formalism we exemplified for our two-scalars model to scenarios previously considered in the literature, starting with the minimal conformal scalar extension of the SM \cite{Allison:2014zya,Allison:2014hna,Sannino:2015wka,Marzola:2016xgb}. The tree level potential is given in this case by\footnote{The Lagrangian we consider admits  a $\mathbb{Z}_2$ symmetry that, being preserved by the symmetry breaking, could lead to the formation of domain walls. The issue is avoided in UV completion of the proposed scenario which explicitly break the $\mathbb{Z}_2$ symmetry or extend it to a continuous  gauge group.}
\begin{equation}
	V(H,s) = \lambda_h (H^\dagger H)^2 + \frac{\lambda_{hs}}{2} (H^\dagger H) s^2 + \frac{\lambda_s}{4} s^4 \,,
\end{equation}
where $H$ is the SM Higgs doublet and $s$ is a real scalar which transforms as a singlet under the gauge symmetry of the SM. Notice that successful electroweak symmetry breaking requires $\lambda_{hs}<0$.

In order to find the direction in the field space where the minimum of the one-loop potential lies, we rewrite the physical Higgs field (in unitary gauge), $h$, and $s$ in polar coordinates $(\phi, \theta)$:
\begin{align}
\begin{cases}
	h = \phi \cos\theta\\
	s=\phi \sin\theta
\end{cases}\,.
\end{align}
Then, we solve for the angle $\theta=\theta^*$ such that ${\rm d}V/{\rm d}\theta\vert_{\theta^*} = 0$, obtaining
\begin{equation}
	\label{eq:flat0}
	\tan^2\theta^* = \frac{2\lambda_h - \lambda_{hs}}{2\lambda_s - \lambda_{hs}} \,,
\end{equation}
and by imposing the condition 
\begin{equation}
	\label{flat}
	\lambda_{hs}^2 - 4\lambda_h \lambda_s = 0 \,,
\end{equation}
the tree-level potential is flat along the $\theta=\theta^*$ direction. We remark that the choice of couplings encoded in eq.~\eqref{flat} is meant to guarantee the existence of a direction in the potential, corresponding to $\theta=\theta^*$, along which quantum corrections dominate over the tree-level contribution. In this way the Coleman-Weinberg mechanism is successfully implemented and the potential acquires the form in eq.~\eqref{eq:CWpot} along such direction. The $B$ parameter is given here by
\begin{equation}
	B = \frac{1}{32\pi^2}\left(W_\sigma^4+3W_Z^4+6W_W^4-12W_t^4\right) \,,
	\label{Bssm}
\end{equation}
where $W_\sigma^2 = -\lambda_{hs}$ is the contribution arising from the tree-level mass of the eigenstate $\sigma$, perpendicular to $\phi$. The remaining quantities depend instead on the SM parameters as follows:
\begin{equation}
	\frac{W_Z^2}{\cos^2\theta^*}  = \frac{g_L^2+g_Y^2}{4}, \quad \frac{W_W^2}{\cos^2\theta^*} = \frac{g_L^2}{4}, \quad \frac{W_t^2}{\cos^2\theta^*} = \frac{y_t^2}{2} \,.
\end{equation}
The expressions for the mass eigenstates in terms of the original fields $h$ and $s$ are given by
\begin{align}
	\begin{cases}
		\phi = h \cos\theta^* + s \sin\theta^*\\
		\sigma = -h \sin\theta^* + s \cos\theta^*
	\end{cases}
\end{align}
whereas the associated masses are
\begin{equation}
	M_\phi^2 = 4Bv_\phi^2\,, \quad M_\sigma^2 = W_\sigma^2v_\phi^2\,,
\end{equation}
with $v_\phi^2=v_h^2+v_s^2 $.

\begin{figure}[h!]
\includegraphics[width=.45\textwidth]{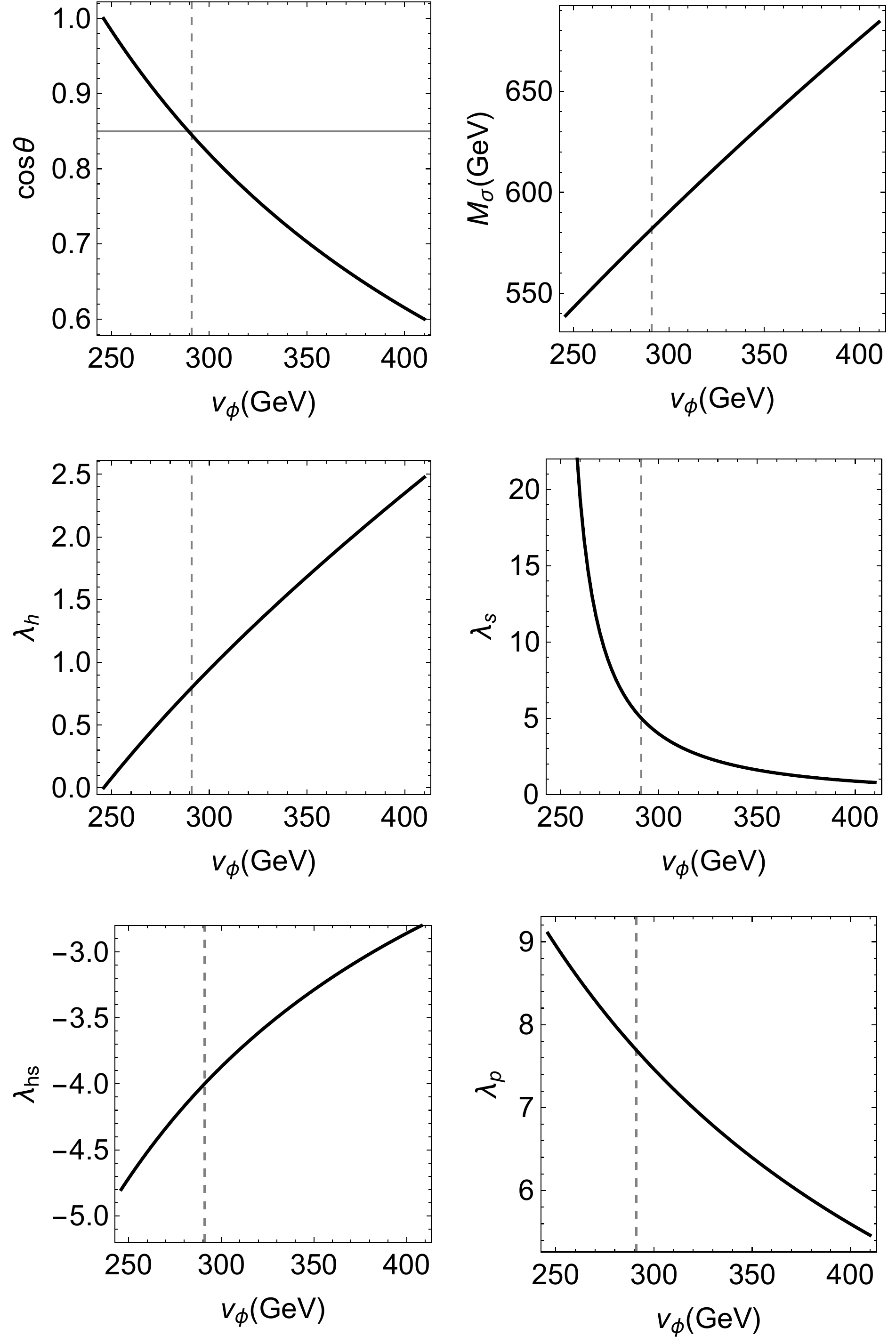}
\caption{Parameters for the scenario I of the minimal conformal SM extension as a function of $v_\phi^2=v_h^2+v_s^2$. Here the scalar boson detected at the LHC corresponds to the flat direction of the tree level potential, which develops a minimum via the Coleman-Weinberg mechanism. The region on the right hand side of the dashed line is excluded by the LHC Higgs phenomenology as it violates the bound $\cos\theta^*>0.85$ \cite{Alanne:2014bra}.}
\label{fig:ssm}
\end{figure}

\subsection*{Scenario I: $ v_s < v_h$ } 
\label{sub:small}

With the above formalism at hand we now investigate the scenario where $\phi$ corresponds to the scalar particle observed at the LHC \cite{Aad:2012tfa,Chatrchyan:2012xdj}. The mixing angle $\theta^*$ is constrained by the LHC Higgs phenomenology to the range  $\cos\theta^*>0.85$ \cite{Alanne:2014bra}, which can be recast through the relation $\tan^2\theta^* = v_s^2/v_h^2$ as an upper bound on the VEV of $s$: $v_s<152$\,GeV. We compute $\lambda_h$ from $M_\phi^2 = M_h^2 = (126\,{\rm GeV})^2$, whereas $\lambda_s$ and  $\lambda_{hs}$ are determined through eq.~\eqref{eq:flat0} and ~\eqref{flat} by using $\tan^2\theta^* = v_s^2/v_h^2$. As for the remaining parameters, we take the Higgs VEV at $v_h=246$\,GeV and the following values for the SM couplings: $g_L=0.648$, $g_Y=0.359$, and $y_t=0.951$. The portal coupling corresponding to $\lambda_p$ in the example of section~\ref{sec:Gravitational signatures of conformal models} is given here by $\lambda_p = \sqrt{2}\,4\pi M_\phi/v_\phi$. 

Fig.~\ref{fig:ssm} shows the parameters of the model as a function of $v_\phi$. We see that in the region allowed by the LHC $\lambda_s$ and $\lambda_{hs}$ assume very large values: $\lambda_s \gtrsim 5$ and $|\lambda_{hs}| \gtrsim 4$.
As a consequence, the scenario is strongly impaired by the presence of a Landau pole at relatively low energies. At best, for the smallest allowed magnitudes of the couplings $\lambda_s \simeq 5$, $\lambda_{hs} \simeq -4$ and $\lambda_h \simeq 0.8$, we estimate with 1-loop RGEs that the Landau pole appears at a scale $\Lambda \sim$ 1 TeV. The consistency of the scenario then imposes the presence of new physics below such scale, clashing with the null results of current LHC searches. 

\subsection*{Scenario II: $v_h < v_s$} 
\label{sub:Large theta scenario}

\begin{figure}[h]
\includegraphics[width=.36\textwidth]{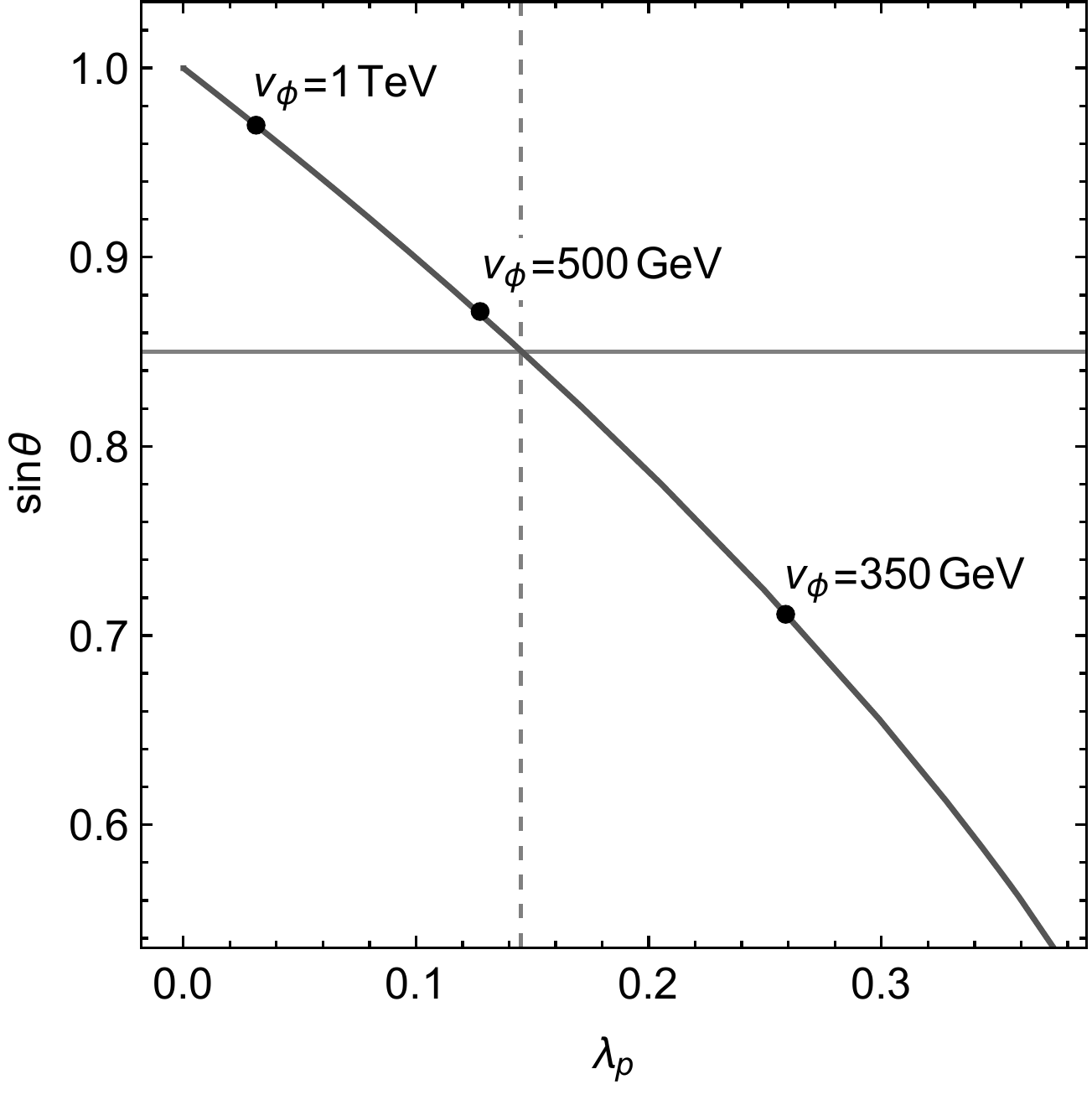}
\caption{The scenario II of the minimal conformal extension of the SM. The observed Higgs boson corresponds here to the direction orthogonal to the flat direction of the tree-level potential. The region on the right hand side of the dashed line is excluded by the LHC Higgs phenomenology. The value of $v_\phi^2=v_h^2+v_s^2$ changes along the solid line as illustrated by the three points.}
\label{fig:ssm2}
\end{figure}

We analyse next the complementary case in which the detected Higgs boson corresponds to $\sigma$. The mixing angle imposed by the LHC Higgs phenomenology is then large, $\sin\theta^*>0.85$, and consequently $v_s>397$\,GeV. For $v_h \ll v_s$ we obtain $\lambda_h \simeq M_h^2/(2 v^2) = 0.131$ for the Higgs quartic coupling, while the $s$ quartic coupling, the portal coupling and the mixing angle are given by:
\begin{equation}
	\lambda_s \simeq \frac{M_h^2 v_h^2}{2v_\phi^4} \,,\quad \lambda_{hs} \simeq -\frac{M_h^2}{v_\phi^2} \,,\quad \sin^2\theta \simeq 1-\frac{v_h^2}{v_\phi^2} \,.
\end{equation}
The portal coupling along the $\phi$ direction matches here $\lambda_p = 2M_h^2/v_\phi^2$ and, in the region consistent with the LHC data, is typically small: $\lambda_p\lesssim0.14$. We plot in fig.~\ref{fig:ssm2} the values obtained for this parameter against the mixing angle. By comparing these results to the ones obtained in section~\ref{sec:Gravitational signatures of conformal models}, we conclude that this scenario is excluded for reheating temperatures high enough to restore the electroweak symmetry via thermal effects. The region of the parameter space associated to the scenario allowed by the LHC constraints leads in fact to a bubble nucleation temperature that violates the lower bound posed by the BBN temperature.

\section{Next-to-minimal model} 
\label{sec:Next-to-minimal models}
We showed in the previous section that the minimal conformal extension of the SM is strongly disfavoured by the electroweak phase transition phenomenology and perturbativity arguments. Here we consider instead the next-to-minimal scenario \cite{Foot:2010av,Ishiwata:2011aa,Farzinnia:2013pga,Khoze:2013uia,Gabrielli:2013hma}, where two new real scalar fields $s$ and $s^\prime$, both singlets under the gauge symmetry of the SM, couple to the Higgs boson. As customary in this framework, we assign a $\mathbb{Z}_2$ symmetry to exclude terms containing odd powers of the new fields in the Lagrangian of the model. The tree level scalar potential then reads
\begin{align}
	V(H,s,s^\prime) =& \lambda_h (H^\dagger H)^2 + \frac{\lambda_{hs}}{2} (H^\dagger H) s^2 + \frac{\lambda_s}{4} s^4 \\\nn +& \frac{\lambda_{hs^\prime}}{2} (H^\dagger H) s^{\prime2} + \frac{\lambda_{ss^\prime}}{4} s^2s^{\prime2} + \frac{\lambda_{s^\prime}}{4} s^{\prime4} \,.
\end{align}
In this analysis we limit ourselves to the case $\lambda_{ss^\prime},\lambda_{s^\prime}>0$, so that $v_\sigma\equiv0$, and set $\lambda_{hs^\prime}=0$ for simplicity. As we will see, this simplified scenario is sufficient to show the potential impact of gravitational wave experiments.   

\begin{figure}[h!]
	\centering
\includegraphics[width=.45\textwidth]{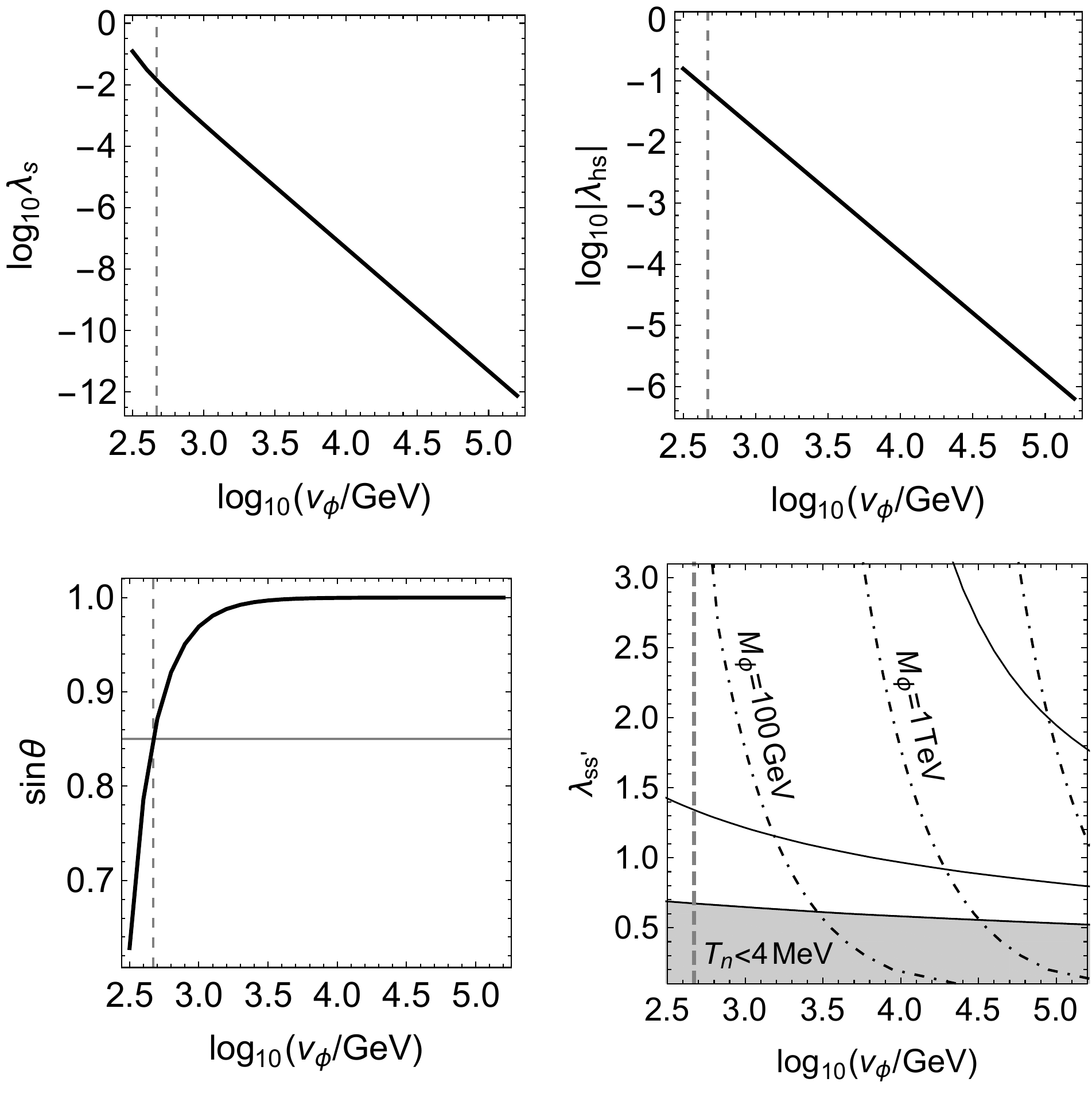}
\caption{Parameters of the next-to-minimal conformal extension of the SM as a function of $v_\phi^2=v_h^2+v_s^2$. The region on the left hand side of the dashed vertical line is excluded by the LHC Higgs boson phenomenology. In the bottom right panel, the solid lines correspond to the $T_n$ contours of fig.~\ref{fig:transition} whereas the dashed contours show the mass of $\phi$.}
\label{fig:sssm}
\end{figure}

In fact, although the presence of a new scalar cannot prevent the Landau pole from appearing at relatively low scale in the setup of the scenario I considered above, the extra portal coupling allows us to recover a correct phase transition dynamics in the complementary case of scenario II. Supposing that $v_s\gg v_h$, the dynamics of this next-to-minimal scenario differs from that of the former case only by the definition of $B$, which includes here the term arising from the portal coupling between $s^\prime$ and $s$: 
\begin{equation}
	W_{s^\prime}^2 = \frac{\lambda_{ss^\prime}}{2} \,.
\end{equation}
The BBN constraint that impairs the scenario II can then be overcome provided that $B$ is dominated by the new contribution above. 

By identifying $\lambda_p \simeq \lambda_{s s^\prime}$ we can discuss the phenomenology of the model by using the results obtained in  section~\ref{sec:Gravitational signatures of conformal models}. In this regard, we plot in fig.~\ref{fig:sssm} the parameters of the model as a function of the VEV $v_\phi$, which as usual lies along the flat direction of the tree-level potential. As we can see, by considering large values of $v_\phi$ we can allow for lower values of $\lambda_p$ (i.e. of $\lambda_{ss^\prime}$), so that the parameters of the model retain perturbativity up to the Grand Unified Theory scale (for $\lambda_{ss^\prime} \lesssim 1.0$) or even Planck scale (in this case $\lambda_{ss^\prime} \lesssim 0.9$). The BBN constraint, instead, is straightforwardly satisfied whenever $\lambda_{s s^\prime}\gtrsim 0.5$ for the reasons previously explained.

As demonstrated in \cite{Gabrielli:2013hma}, in absence of effects that break the $\mathbb{Z}_2$ symmetry imposed on the Lagrangian, $s^\prime$ is a stable particle which can play the role of DM candidate. A first rough estimate of the the relative relic abundance shows that the observed value can be matched through the freeze-out mechanism\footnote{For $ \lambda_{ss^\prime}\sim\mathcal{O}(1)$ the dark matter candidate  thermalises in the early Universe provided that $\lambda_{hs}\gtrsim 10^{-7}$.} via annihilations of $s^\prime$ to the $ss$ final state for a range of values of the involved parameters. A more careful assessment of the DM relic density in the scenario and the detailed gravitational wave phenomenology of the considered next-to-minimal conformal extension of the SM will be presented in a forthcoming analysis.

\section{Conclusions} 
\label{sec:conclusions}
In this paper we studied the phase transition dynamics of scalar classically conformal scenarios posing particular attention to their possible gravitational wave signatures. After having reviewed the basis of the framework and showed that it necessarily leads to a first order phase transition at finite temperature, we studied the properties of a general two-scalars model that captures the gist of the conformal extensions of the Standard Model. We find that the phase transition is generally very strong and leads to the production of a stochastic gravitational wave background which can be observed in current and next-generation dedicated experiments.

We then applied the analysis for the minimal classically conformal scalar  extension of the Standard Model, where the presence of a new scalar field coupled to the Higgs boson implements a Coleman-Weinberg type of potential. The scenario can be analysed in two limits, depending on the hierarchy between the vacuum expectation values of the involved scalar fields. In the case where the vacuum expectation value of the singlet scalar field is smaller than that of the Higgs field, the quartic couplings of the model are very large and result in a Landau pole at the TeV scale. Although the problem could be solved by invoking the presence of new physics, the null result from the LHC disfavour this possibility.

In the complementary regime, where the vacuum expectation value of the new scalar boson is larger than the Higgs field one, we find that the model retains its perturbativity up to the Planck scale. In spite of that, the smallness of the portal coupling between the two scalars delays the electroweak phase transition to temperatures below the Big Bang Nucleosynthesis one, generally excluding the scenario. 

Lastly, we considered a next-to-minimal conformal extension of the Standard Model involving an additional scalar field. In this case we showed that the presence of such particle allows to satisfy the constraint from Big Bang Nucleosynthesis and perturbativity, which impaired the minimal extension. We find that, in the considered regime, the next-to-minimal extension accomplishes a first order electroweak phase transition which gives rise to a sizeable gravitational wave signal, demonstrating the capability of dedicated experiments to explore the scenario.

\section*{Acknowledgements}
This work was supported by the Estonian Research Council grant IUT23-6, PUT1026 and ERDF Centre of Excellence project TK133.

\bibliography{conf}
\end{document}